\documentclass[10pt,journal]{IEEEtran}

\ifCLASSINFOpdf
\else
   \usepackage[dvips]{graphicx}
\fi
\usepackage{url,cite}
\usepackage{multirow}
\usepackage{algorithmic}
\usepackage{algorithm}
\usepackage{amsthm}
\newtheorem{theorem}{Theorem}

\newtheorem{corollary}{Corollary}
\usepackage{amssymb}
\usepackage{amsmath}
\usepackage{amsmath, amssymb, amsthm}
\usepackage{amsmath,amsfonts}
\usepackage{textcomp}
\usepackage{stfloats}
\usepackage{url}
\usepackage{verbatim}
\usepackage{graphicx}
\usepackage{cite}
\usepackage{bm}
\usepackage{stmaryrd}
\usepackage{booktabs}
\usepackage{balance}
\usepackage{siunitx}
\usepackage{xcolor}
\definecolor{deepred}{RGB}{122,1,1}
\usepackage{makecell}
\usepackage{array} 
\newtheorem*{Rem}{Remark}

\theoremstyle{definition}

\usepackage{graphicx}

\begin{document}

\title{Bit-Efficient Quantisation for Two-Channel Modulo-Sampling Systems}

\author{Wenyi Yan, Zeyuan Li, Lu Gan, Honqing Liu, Guoquan Li
\thanks{
This work has been submitted to the IEEE for possible publication. Copyright may be transferred without notice, after which this version may no longer be accessible. }
\thanks{}}

\markboth{Journal of \LaTeX\ Class Files, Vol. 14, No. 8, August 2015}
{Shell \MakeLowercase{\textit{et al.}}: Bare Demo of IEEEtran.cls for IEEE Journals}
\maketitle

\begin{abstract}
Two-channel modulo analog-to-digital converters (ADCs) enable high-dynamic-range signal sensing at the Nyquist rate per channel, but existing designs quantise both channel outputs independently, incurring redundant bitrate costs.
This paper proposes a bit-efficient quantisation scheme that exploits the integer-valued structure of inter-channel differences, transmitting one quantised channel output together with a compact difference index.
We prove that this approach requires only 1–2 bits per signal sample overhead relative to conventional ADCs, despite operating with a much smaller per-channel dynamic range.
Simulations confirm the theoretical error bounds and bitrate analysis, while hardware experiments demonstrate substantial bitrate savings compared with existing modulo sampling schemes, while maintaining comparable reconstruction accuracy.
These results highlight a practical path towards high-resolution, bandwidth-efficient modulo ADCs for bitrate-constrained systems.
\end{abstract}


\IEEEpeerreviewmaketitle

\section{Introduction}
\label{sec:intro}
The unlimited sensing framework (USF) prevents clipping by folding high-amplitude signals into a low dynamic range, enabling digital recovery of signals that would otherwise saturate~\cite{bhandari_unlimited_2021,bhandari_unlimited_2022,bhandari_unlimited_2018,GalUnlimited,ABSparse,DorianHys,DoranTimeEnco,ShtHDR,Guo2023_ICASSP_ITERSIS,Guo6K4,wang2025line}.
Several hardware prototypes have further demonstrated the potential of single-channel USF systems in practice~\cite{bhandari_unlimited_2022,ThomasRadar,mulleti_hardware_2023,zhu2025ironing,ZhuHard}.
Although a variety of algorithms and hardware implementations exist, single-channel USF systems typically require high oversampling rates and complex reconstruction procedures, limiting their suitability for bitrate-constrained or low-cost receivers.

Multi-channel architectures~\cite{LuMulti,gong_multi-channel_2021,ICASSP2024Bits,ICASSP2025,yan_parameter_2025} based on the robust Chinese Remainder Theorem (RCRT)~\cite{WangRCRT,xiaoCRT1,xiaoCRT2} offer a compelling alternative. 
By employing channels with distinct dynamic ranges, they achieve high overall range without per-channel oversampling
~\cite{WangRCRT,gong_multi-channel_2021,ICASSP2024Bits}.
Beyond RCRT, Guo~\emph{et al.}~\cite{guo2023unlimited} proved perfect recovery for finite-rate-of-innovation (FRI) signals under irrational threshold ratios in two-channel modulo sampling systems. However, such a ratio is impractical in hardware and its recovery relies on exhaustive search over wrapping indices.
Several recent works have studied sub-Nyquist architectures for complex sinusoidal mixtures~\cite{ZhuSub,Guo6K4,PVMulti} by employing multicoset sampling with amplitude-unfolding and spectral estimation. Besides, Wang~\emph{et al.}~\cite{wangestimating} formulate sinusoidal estimation as a mixed-integer program using first-order differences and a multi-channel ``virtual'' modulo ADC to enlarge the effective dynamic range. Florescu~\cite{DoranMulti} proposes a multi-channel architecture with a shared folding loop to improve noise robustness via averaging.

A key challenge in multi-channel modulo-ADC systems is quantisation
efficiency.  In existing designs, each channel is sampled and quantised
independently using $b$ bits per modulo sample.  In particular,~\cite{ICASSP2024Bits} and~\cite{PVMulti} provide
closed-form expressions for the minimum required $b$ in systems with 
 $L$ real-valued moduli and two complex-valued moduli, respectively.
The total bitrates can be much higher than those of a
conventional ADC. To overcome this limitation, we develop an Efficient
CRT (ECRT) quantisation scheme that exploits inter-channel redundancy
within a two-channel modulo sampling architecture.
Our main contributions are as follows:
\begin{itemize}
    \item \textbf{Bit-efficient quantisation:} The proposed ECRT transmits the quantised modulo samples from \emph{one} channel together with a compact quantised inter-channel difference, eliminating redundancy and substantially reducing the total bits per signal sample compared with existing two-channel systems~\cite{ICASSP2024Bits}.
    
    \item \textbf{Theoretical guarantees:} Closed-form bounds on the recovery error and bit requirements are derived, showing that ECRT requires only a 1–2 bit per sample overhead relative to a conventional high-dynamic-range ADC, with a single-bit overhead in most cases.
    
    \item \textbf{Simulation and hardware validation:} Simulations confirm the theoretical error bounds and bitrate predictions, while hardware experiments demonstrate that ECRT achieves a significantly lower bitrate than single-channel modulo sampling schemes, offering a more communication-efficient solution.
\end{itemize}

\section{Background and Motivation}\label{sec:background}

For an input $x \in \mathbb{R}$ and ADC dynamic range $\Delta > 0$, the modulo operation is defined as~\cite{bhandari_unlimited_2018}
\begin{equation}\label{eq:mod-def}
\left\langle x \right\rangle_\Delta 
= x - \Delta \left\lfloor \frac{x}{\Delta} + \tfrac{1}{2} \right\rfloor,
\end{equation}
which folds $x$ into $\left[-\tfrac{\Delta}{2}, \tfrac{\Delta}{2}\right)$.
Consider a bandlimited signal $g(t)$ processed by a two-channel modulo-ADC system with 
\begin{equation}\label{eq:defdelta}
    \Delta_\ell = \tau_\ell \varepsilon, \quad \ell=1,2,
\end{equation}
where $\tau_1, \tau_2$ are coprime integers satisfying $2 \leq \tau_1 < \tau_2$, and $\varepsilon > 0$ is a scaling factor. Sampling at period $T$ yields
\begin{equation}\label{eq:tildeyk}
    \tilde{y}_\ell[k] = \left\langle g(kT) \right\rangle_{\Delta_\ell} + e_\ell[k]=y_\ell[k]+ e_\ell[k], \quad k \in \mathbb{Z},
\end{equation}
where $y_\ell[k] = \left\langle g[k] \right\rangle_{\Delta_\ell}$ is the noiseless modulo output and $e_\ell[k]$ is the folding noise. $g[k]$ can be written as
\begin{equation}\label{eq:defgk}
    g[k] = \Delta_\ell n_\ell[k] + y_\ell[k],
\end{equation}
where $n_\ell[k] \in \mathbb{Z}$ are folding indices. 
Quantising each $\tilde y_\ell[k]$ with $b$ bits gives
\begin{equation}\label{eq:hatyk}
    \hat y_\ell[k]=\tilde{y}_\ell[k]+q_\ell[k] =y_\ell[k]+ e_\ell[k]+q_\ell[k], \quad k \in \mathbb{Z},
\end{equation}
in which the quantisation noise $q_\ell[k]$ is bounded by $|q_\ell[k]|\le \Delta_\ell/2^{\,b+1}$. RCRT estimates folding indices from $\hat y_1[k]-\hat y_2[k]$ and reconstructs via~\cite{WangRCRT}
\begin{equation}\label{eq:crt-gkexp}
\hat g_{\mathrm{RCRT}}[k] = \tfrac{1}{2}\Big( \hat n_1[k]\Delta_1 + \hat y_1[k] 
+ \hat n_2[k]\Delta_2 + \hat y_2[k]\Big).
\end{equation}
Stable recovery ($\hat n_\ell[k]=n_\ell[k]$) is guaranteed when~\cite{WangRCRT}
\begin{equation}\label{eq:rcrt-cond}
    |e_2[k]-e_1[k]| + \frac{\Delta_1+\Delta_2}{2^{\,b+1}} < \frac{\varepsilon}{2}
\end{equation}
The bit-depth requirement for stable recovery is~\cite{ICASSP2024Bits} 
\begin{equation}\label{eq:samplingBits}
B_{\mathrm{RCRT}} = 2b, \quad \text{with} \quad b \ge \left\lceil \log_2(\tau_1+\tau_2)\right\rceil,
\end{equation}
where $B_{\mathrm{RCRT}}$ is the total number of bits per sample, and equality holds when folding noise is absent ($e_\ell[k]=0$).

For comparison, define the amplitude scaling factor
\begin{equation}\label{eq:rho-def}
\rho = 2\|g(t)\|_\infty/\Delta,
\end{equation}
and let $B_c$ denote the bit cost of a conventional ADC (same quantisation step). Then~\cite{ICASSP2024Bits}
\begin{equation}\label{eq:con}
B_c = b + \left\lceil \log_2 \rho \right\rceil .
\end{equation}
For a given $\rho$, we need $\tau_1 \ge \rho$ and $\tau_2 \ge \rho+1$~\cite{ICASSP2024Bits}, yielding $b \ge \lceil \log_2(\rho+1) + 1 \rceil$ and an overhead
\begin{equation}\label{eq:ext}
B_{\text{RCRT}} - B_c = b - \left\lceil \log_2 \rho \right\rceil,
\end{equation}
that grows linearly with $b$. We next introduce a quantisation scheme reducing this overhead to only 1 to 2 bits.

\section{Proposed Bit-Efficient Two-Channel Design}
\label{sec:bit_efficient}

\begin{figure}
    \centering
    \includegraphics[width=0.8\linewidth]{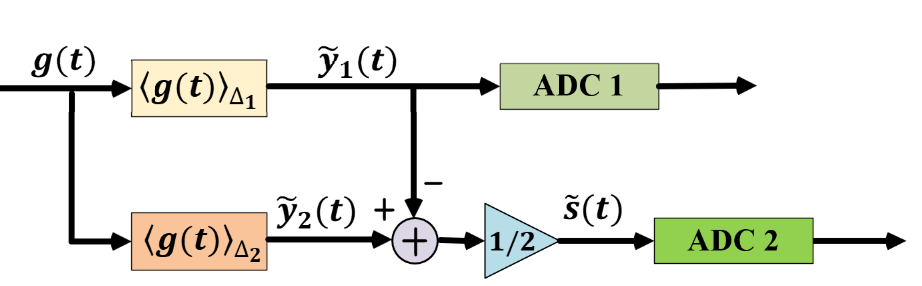}
\caption{Proposed bit-efficient two-channel modulo-ADC architecture.
Channel~1 quantises $y_1(t)$, while Channel~2 quantises the scaled signal $\tilde{s}(t)$.}
    \label{fig:archi}
\end{figure}

\subsection{System Model}

To obtain a bit-efficient scheme, we first exploit the inter-channel redundancy. Define the
normalised difference
$d[k] = \frac{y_2[k]-y_1[k]}{\varepsilon}$.
From~\eqref{eq:defdelta} and~\eqref{eq:defgk},
\[
d[k]
= \frac{y_2[k]-y_1[k]}{\varepsilon}
= \tau_1 n_1[k] - \tau_2 n_2[k],
\]
so $d[k]\in\mathbb{Z}$. Using $-\Delta_\ell/2 \le y_\ell[k] <
\Delta_\ell/2$, we obtain $|d[k]|<(\tau_1+\tau_2)/2$, showing that
$d[k]$ takes at most $\tau_1+\tau_2$ integer values. 
However, directly forming the difference $y_2[k]-y_1[k]$ expands the
dynamic range to $(\Delta_1+\Delta_2)$, which exceeds the range of
either ADC. To avoid this saturation issue, we instead sample the
scaled difference
$s[k]=\frac{y_2[k]-y_1[k]}{2}$,
for which $|s[k]|<(\Delta_1+\Delta_2)/2<\Delta_2/2$. Thus $s[k]$
always lies strictly within the dynamic range of the second ADC.

These observations motivate the proposed two-channel modulo sampling
architecture in Fig.~\ref{fig:archi} and the associated quantisation
and recovery scheme in Alg.~\ref{alg:bitCRT}. Theorem~\ref{thm:ecrt-final} quantifies the bit budget required by the
proposed design and establishes its stability under bounded noise.

\begin{theorem}[ECRT: quantisation and stable recovery]
\label{thm:ecrt-final}
Consider the two-channel modulo sampling system in Fig.~\ref{fig:archi} with
dynamic ranges $\Delta_\ell=\tau_\ell\varepsilon$, $\ell=1,2$, where
$\varepsilon>0$ and $\tau_1,\tau_2$ are relatively coprime integers
satisfying $2\le\tau_1<\tau_2$.  For an input signal $g(t)$ with $\|g(t)\|_\infty\le \frac{\tau_1\tau_2\varepsilon}{2}$ sampled at
period $T$, let $y_\ell[k]=\langle g(kT)\rangle_{\Delta_\ell}$ denote
the clean modulo samples, $\tilde y_1[k]=y_1[k]+e_1[k]$ the noisy
samples at channel~1, and 
\[
\tilde s[k]=\varepsilon d[k]/2 + e_s[k]
\]
the sampled difference signal, where
$d[k]=(y_2[k]-y_1[k])/\varepsilon$ is the normalised difference.
Assume that $\tilde y_1[k]$ and $\tilde s[k]$ are quantised with $b\ge1$
and $b_d$ bits, respectively as in Alg.~\ref{alg:bitCRT}, giving a total bit
cost
\begin{equation}\label{eq:bitecrt}
B_{\mathrm{ECRT}} = b + b_d, \qquad 
b_d = \lceil \log_2(\tau_1+\tau_2) \rceil.
\end{equation}
If the difference-path noise satisfies $|e_s[k]| < \varepsilon/4$, 
then the reconstruction error satisfies
\begin{equation}\label{eq:rec-error-ecrt}
|\hat g_{\mathrm{ECRT}}[k] - g[k]| 
\le \|e_1\|_\infty + \frac{\Delta_1}{2^{b+1}},
\end{equation}
where $\|e_1\|_\infty=\max_k |e_1[k]|$.
\end{theorem}
\begin{IEEEproof}
From the quantisation of $\tilde s[k]$ with step $\varepsilon/2$,
the dequantised value satisfies
$\hat s[k]=\tilde s[k]+q_s[k]$ with $|q_s[k]|\le\varepsilon/4$.
Under the noise condition $|e_s[k]|<\varepsilon/4$, we have
\[
|\hat s[k]-\tfrac{\varepsilon d[k]}{2}|
\le |e_s[k]| + |q_s[k]| < \tfrac{\varepsilon}{2},
\]
which ensures
$\hat d[k]=\mathrm{round}(2\hat s[k]/\varepsilon)=d[k]$.
Hence the folding index of channel~1 is recovered exactly as
$\hat n_1[k]=(\hat d[k]\gamma_1)\bmod\tau_2=n_1[k]$, where $\gamma_{1}$ denotes the multiplicative inverse of $\tau_{1}$ modulo $\tau_{2}$, i.e.,
$\gamma_{1}\tau_{1} \equiv 1 \pmod{\tau_{2}}$.

The reconstructed signal can be expressed as
\[
\hat g[k]
= \hat n_1[k]\Delta_1 + \hat y_1[k]
= n_1[k]\Delta_1 + (y_1[k]+e_1[k]+q_1[k]),
\]
where $|q_1[k]|\le \Delta_1/2^{b+1}$ is the channel-1 quantisation
error. As $g[k]=n_1[k]\Delta_1+y_1[k]$, we obtain
\[
|\hat g[k]-g[k]|
\le |e_1[k]| + |q_1[k]|
\le \|e_1\|_\infty + \frac{\Delta_1}{2^{b+1}},
\]
which proves~\eqref{eq:rec-error-ecrt}. The bit cost~\eqref{eq:bitecrt}
follows from Alg.~\ref{alg:bitCRT}.
\end{IEEEproof}

\begin{algorithm}[t]
\caption{ECRT Scheme}
\label{alg:bitCRT}
\begin{algorithmic}[1]
\REQUIRE Noisy modulo output $\tilde y_1[k]$; difference signal $\tilde s[k]$; ADC parameters $\Delta_\ell,\:\tau_\ell,\:\varepsilon$ ($\ell=1,2$); bit budget $b$
\ENSURE Reconstructed sample $\hat g[k]$
\item[] \textbf{Transmitter:}
\STATE Quantize $\tilde y_1(t)$ with step $\Delta_1/2^b$ (using $b$ bits)
\STATE $b_d=\lceil\log_2(\tau_1+\tau_2)\rceil$
\STATE Quantize $\tilde s(t)$ with step $\varepsilon/2$ (using $b_d$ bits)
\item[] \textbf{Receiver:}
\STATE Dequantize to obtain $\hat y_1[k]$ and $\hat s[k]$
\STATE $\hat d[k]=\mathrm{round}(2\hat s[k]/\varepsilon)$
\STATE $\hat n_1[k]=(\hat d[k]\gamma_1)\bmod \tau_2$, where $\gamma_1\tau_1\equiv1~(\mathrm{mod}~\tau_2)$
\STATE $\hat g[k]=\hat n_1[k]\Delta_1+\hat y_1[k]$
\end{algorithmic}
\end{algorithm}

\subsection{Comparison with other schemes}

\textbf{Comparison with RCRT~\cite{ICASSP2024Bits}.} 
ECRT requires a fixed $b_d$ bits to quantise $\tilde{s}(t)$,
and channel~1 may use as little as $b=1$ bit. In contrast, RCRT requires $b$ to satisfy
\eqref{eq:bitecrt}. Under this condition, combining \eqref{eq:samplingBits} and
\eqref{eq:bitecrt} gives $B_{\mathrm{RCRT}} - B_{\mathrm{ECRT}} \ge 0$, showing that
ECRT always achieves a smaller bit budget.
In practice, forming the scaled
difference $\tilde s[t]$ incurs negligible analog cost, as
subtraction and scaling of $1/2$ are readily implemented by a simple
operational amplifier. The noise term $e_s[k]$ is dominated by the
modulo-stage errors, and the ECRT condition $|e_s[k]|<\varepsilon/4$ is
effectively equivalent to the RCRT requirement
$|e_2[k]-e_1[k]|<\varepsilon/2$, indicating that both schemes exhibit
comparable noise tolerance.

For reconstruction error, when $\hat n_\ell[k]=n_\ell[k]$ for
$\ell=1,2$, RCRT's averaging gives
\[
\hat g_{\mathrm{RCRT}}[k]-g[k]
= \tfrac{1}{2}\big[(e_1[k]+q_1[k])+(e_2[k]+q_2[k])\big].
\]
With $\|e\|_\infty \triangleq \max_{k,\ell} |e_\ell[k]|$, this yields
\[
|\hat g_{\mathrm{RCRT}}[k]-g[k]|
\le \|e\|_\infty + \frac{\Delta_1+\Delta_2}{2^{b+2}}.
\]

Compared with~\eqref{eq:rec-error-ecrt}, both schemes achieve the same worst-case error order. 
For average-case performance, the key distinction lies in how channel outputs are combined. ECRT reconstructs solely from the first-channel residue $\hat y_1[k]=y_1[k]+e_1[k]+q_1[k]$, yielding the error $e_1[k]+q_1[k]$. In contrast, RCRT averages both residues, and the
reconstruction error is
\[
\hat g_{\mathrm{RCRT}}[k]-g[k] = \tfrac{1}{2}\big( e_1[k]+q_1[k]
                              +e_2[k]+q_2[k] \big).
\]
Under independent zero-mean noise,
\[
\mathrm{Var}\left(\hat g_{\mathrm{RCRT}}[k]-g[k]\right)
   = \tfrac{1}{4}\big( \mathrm{Var}(e_1+q_1)
                     + \mathrm{Var}(e_2+q_2) \big)
\]
which reduces to around $\tfrac{1}{2}\,\mathrm{Var}(e_1+q_1)$ when $\Delta_1\approx\Delta_2$. Thus, RCRT nearly halves noise variance through averaging, while ECRT trades this reduction for lower bitrate.

\textbf{Comparison with conventional ADC.}
We compare the ECRT against a conventional high-resolution ADC without saturation. 
For ECRT, the moduli $\tau_1=\lceil\rho\rceil,\ \tau_2=\lceil\rho\rceil+1$ are chosen to maximize the error tolerance parameter $\varepsilon$ for a given $\rho$~\cite{ICASSP2024Bits}, where $\lceil\cdot\rceil$ denotes the ceil operation.
\begin{corollary}
\label{cor:ecrt-vs-conv}
Let $\rho>1$ be defined in~\eqref{eq:rho-def}. With $B_c$ from~\eqref{eq:con} and $B_{\mathrm{ECRT}}$ from~\eqref{eq:bitecrt} using $\tau_1=\lceil\rho\rceil,\ \tau_2=\lceil\rho\rceil+1$, for any integer $n\ge1$,
\[
B_{\mathrm{ECRT}}=
\begin{cases}
B_c+1, & 2^{n-1}<\rho\le 2^n-1,\\
B_c+2, & 2^n-1<\rho\le 2^n.
\end{cases}
\]
\end{corollary}

\begin{IEEEproof}
If $2^{n-1}<\rho\le2^n$, then $\lceil\log_2\rho\rceil=n$ so $B_c=b + n$. For $2^{n-1}<\rho\le2^n-1$, one has $2^{n-1}+1\le\lceil\rho\rceil\le2^n-1$, hence $2^n+3<2\lceil\rho\rceil+1\le2^{n+1}-1$, giving $\lceil\log_2(2\lceil\rho\rceil+1)\rceil=n+1$ and $B_{\mathrm{ECRT}}=B_c+1$. If $2^n-1<\rho\le2^n$, then $\lceil\rho\rceil=2^n$ and $2^{n+1}<2\lceil\rho\rceil+1\le2^{n+1}+1$, yielding $\lceil\log_2(2\lceil\rho\rceil+1)\rceil=n+2$ and $B_{\mathrm{ECRT}}=B_c+2$.
\end{IEEEproof}

\begin{Rem}
A two-bit overhead arises only on the short interval $(2^n-1,2^n]$ (length $1$) inside the dyadic range $(2^{n-1},2^n]$ (length $2^{n-1}$). Hence, ECRT typically requires only one extra bit per signal sample compared with a conventional ADC, even though it operates with a much smaller dynamic range. 
\end{Rem}

\begin{table}[t]
\centering
\setlength{\tabcolsep}{8pt}
\renewcommand{\arraystretch}{1.2}
\caption{Theoretical bitrates (bps) of conventional ADC, SOSI~\cite{bernardo_modulo_2024} and ECRT 
for signal bandwidth $0.5$~Hz (Nyquist rate $1$~Hz).}
\label{tab:rate_comparison}
\begin{tabular}{c|c|c|c|c}
\toprule
\hline
$b$ & $\rho$ & $R_{\text{C}}$ & $R_{\text{SOSI}}$ & $R_{\text{ECRT}}$ \\
\hline
\multirow{3}{*}{3} 
 & 5  & 6  & 28 & 7  \\ \cline{2-5}
 & 8  & 6  & 40 & 8  \\ \cline{2-5}
 & 10 & 7  & 48 & 8  \\ 
\hline
\multirow{3}{*}{6} 
 & 5  & 9  & 49 & 10 \\ \cline{2-5}
 & 8  & 9  & 70 & 11 \\ \cline{2-5}
 & 10 & 10 & 84 & 11 \\
 \hline
\bottomrule
\end{tabular}
\end{table}

\textbf{Comparison with single-channel systems.} 
For a bandlimited signal $g(t)$ with finite energy, we benchmark the proposed design against single-channel modulo sampling with one-bit side information (SOSI)~\cite{bernardo_modulo_2024,SOSI}, which provides closed-form guarantees on oversampling factor $\mathrm{OF}\triangleq f_s/f_{\mathrm{NYQ}}$ and stable recovery.
From~\cite[Lem.~1 \& Thm.~1]{bernardo_modulo_2024}, SOSI requires 
$b > 3$, $\mathrm{OF} > 3$, and dynamic range $\rho \le \mathrm{OF} - 2$, 
yielding bit rate
\[
R_{\mathrm{SOSI}} = (b+1)\,\mathrm{OF}\,f_{\mathrm{NYQ}}
\ge (b+1)(\rho+2)\,f_{\mathrm{NYQ}},
\]
which scales \emph{linearly} with $\rho$ and incurs $\mathcal{O}(N^3)$ reconstruction complexity for $N$ signal samples.
In contrast, ECRT operates at Nyquist rate ($\mathrm{OF}=1$) with $\mathcal{O}(N)$ complexity and bit rate
\[
R_{\mathrm{ECRT}} = \big(b + \lceil \log_2(2\rho+1) \rceil \big)\,f_{\mathrm{NYQ}},
\]
where $b = \Theta(\log \rho)$, yielding only \emph{logarithmic} scaling in $\rho$. This makes ECRT substantially more efficient in bitrate-constrained scenarios, achieving comparable error to conventional ADCs without oversampling or complex decoding.

While these architectures represent different design trade-offs (oversampling versus multi-channel redundancy), ECRT is particularly advantageous for bandwidth-limited applications.

\begin{Rem}
Under high oversampling, SOSI attains an MSE scaling of
$\mathcal{O}(\mathrm{OF}^{-3})$~\cite[Thm.~1]{bernardo_modulo_2024},
surpassing the $\mathcal{O}(\mathrm{OF}^{-1})$ decay of conventional
oversampled ADCs, but only when $\mathrm{OF}>\rho+2$ and at the cost of
increased bitrate and decoder complexity. Hardware results
(Section~\ref{sec:simulation}) show that although SOSI offers slightly
better accuracy at large oversampling factors, ECRT achieves comparable
performance at substantially lower bitrate, making it preferable when
communication bandwidth or receiver resources are constrained.
\end{Rem}

\textbf{Sub-Nyquist Extension:} 
For spectrally sparse signals, the proposed framework can be combined with multicoset sampling~\cite{multicoset1,mutlcoset2} or coprime sensing~\cite{xiaoCRT4,xiaCRT} to further reduce sampling rate.
For example, introducing a delayed coset for each modulo channel (four branches total)~\cite{Guo6K4} enables sampling periods $T > 1/f_{\mathrm{NYQ}}$ while preserving sub-Nyquist spectral estimation. In this setting, ECRT performs quantisation and amplitude unfolding across each coset pair, providing bitrate-efficient amplitude unfolding. We leave this extension to future work.

\section{Simulation and Hardware Results}
\label{sec:simulation}

\textbf{Simulation.} We compare ECRT with conventional ADC and RCRT-based system~\cite{ICASSP2024Bits} for a bandlimited input of bandwidth $0.5$~Hz ($f_{\mathrm{NYQ}}=1$~Hz) and peak amplitude $\|g(t)\|_\infty=22000$. 
With $\Delta_2=300$, the amplitude scaling factor is $\rho=2\|g(t)\|_\infty/\Delta_2=14.67$. Following~\cite{ICASSP2024Bits}, the optimal two-channel setting uses consecutive coprimes $\tau_1=\lceil\rho\rceil=15$ and $\tau_2=16$, giving $\varepsilon=\Delta_2/\tau_2\approx18.75$ and thresholds $\Delta_1\approx281.25,\ \Delta_2=300$. 
Performance is measured by the maximum absolute error (MAE) \[E=\max_k|g[k]-\hat g[k]|.\]

\begin{figure}[t]
    \centering
    \includegraphics[width=0.7\linewidth]{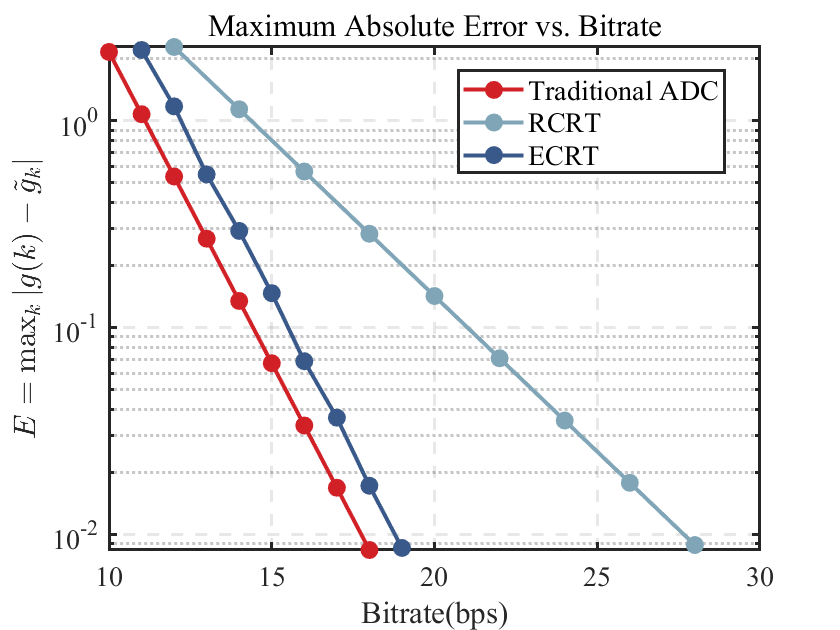}
\caption{MAE vs. transmission bitrate for conventional ADC, RCRT, and the proposed ECRT. }
    \label{fig:CRT}
    \vspace{-0.4cm}
\end{figure}

Fig.~\ref{fig:CRT} shows MAE versus bitrate. As expected, the conventional ADC provides the benchmark, while ECRT closely follows with only a minor offset. In contrast, RCRT incurs higher redundancy since both channels are transmitted. For example, to achieve MAE $\approx10^{-2}$, conventional ADC requires about $18$~bps, ECRT $19$~bps, whereas RCRT needs nearly $28$~bps. These results are consistent with Corollary~\ref{cor:ecrt-vs-conv}, confirming that ECRT introduces only a one-bit overhead per sample compared with a conventional ADC.

\begin{table*}[htb]
\centering
\caption{Bitrate (kbps) and RRSE for a bandlimited input with $\rho=2.8$ and $B=\SI{10}{\kilo\hertz}$ under different quantisation schemes.}
\renewcommand{\arraystretch}{1.3}
\begin{tabular}{c|cc|cc|cc|cc}
\toprule
\hline
& \multicolumn{2}{c|}{US-ALG~\cite{bhandari_unlimited_2021}} 
& \multicolumn{2}{c|}{USLSE~\cite{ZhuLSE}} 
& \multicolumn{2}{c|}{SOSI~\cite{bernardo_modulo_2024}} 
& \multicolumn{2}{c}{Proposed ECRT} \\
\cline{2-9}
$b$& \multicolumn{2}{c|}{$f_s=73.53$ kHz} 
& \multicolumn{2}{c|}{$f_s=119.05$ kHz} 
& \multicolumn{2}{c|}{$b_e=1$, $f_s=51.02$ kHz} 
& \multicolumn{2}{c}{$b_d=3$, $f_s=20$ kHz} \\
\cline{2-9}
& Bitrate & RRSE  & Bitrate & RRSE & Bitrate & RRSE & Bitrate & RRSE \\
\hline
3 & 220.59  & 15.47 & 357.14  & $9.17\times 10^{-2}$ & 204.08 & $7.21\times 10^{-2}$ & 120.00 & $7.23\times 10^{-2}$ \\ 
\hline
4 & 294.12 & $4.52\times 10^{-2}$  & 476.19  & $7.22\times 10^{-2}$ & 255.10 & $4.95\times 10^{-2}$ & 140.00 & $5.37\times 10^{-2}$ \\ 
\hline
5 & 367.65  & $3.62\times 10^{-2}$  & 595.24  & $6.53\times 10^{-2}$ & 306.12 & $3.99\times 10^{-2}$ & 160.00 & $4.93\times 10^{-2}$ \\ 
\hline
6 & 441.18  & $3.17\times 10^{-2}$  & 714.29  & $6.22\times 10^{-2}$ & 357.14 & $3.87\times 10^{-2}$ & 180.00 & $4.81\times 10^{-2}$ \\ 
\hline
\bottomrule
\end{tabular}
\label{tab:comparison_rrse}
\end{table*}

\textbf{Hardware Experiment.} We validate the proposed ECRT-based two-channel modulo ADC through experimental measurements obtained from our hardware prototype~\cite{MultiADCsISCAS}.
The setup uses $\Delta_1=0.72~\text{V}$ and $\Delta_2=0.96~\text{V}$ (i.e., $\tau_1=3$, $\tau_2=4$ and $\varepsilon=0.24$) to acquire a $\SI{10}{\kilo\hertz}$ bandlimited signal ($\rho=2.8$). 
Although our FPGA prototype supports real-time recovery using its
on-board ADC~\cite{MultiADCsISCAS}, all hardware analog measurements in this section are acquired
using an 8-bit Tektronix TDS\,1012C-EDU oscilloscope. This ensures a
fair comparison with single-channel systems: the oscilloscope is
used to capture the ground-truth signal, the analog modulo outputs, and
the analog difference signal, thereby avoiding additional scaling or
timing jitter introduced by the on-board ADC. The reconstruction
accuracy is then evaluated using
the relative root-squared error (RRSE), defined as 
\[
\mathrm{RRSE}=\sqrt{\frac{\sum_k |g[k]-\hat g[k]|^2}{\sum_k |g[k]|^2}}.
\]

Table~\ref{tab:comparison_rrse} compares the proposed ECRT with representative single-channel methods: US-ALG~\cite{bhandari_unlimited_2021}, USLSE~\cite{ZhuLSE}, and SOSI~\cite{bernardo_modulo_2024}. 
ECRT quantises the scaled difference index with $b_d=3$ bits, while SOSI uses one extra bit ($b_e=1$). 
The results show that ECRT achieves comparable RRSE to the single-channel schemes, which require $f_s > 50\ \text{kHz}$ for similar fidelity, while operating at only $20\ \text{kHz}$. 
Across bit depths $b=3$–$6$, ECRT achieves substantial bitrate reduction relative to SOSI, with negligible accuracy loss.     

The theoretical analysis (Table~\ref{tab:rate_comparison}) predicts that ECRT's bitrate savings grow with $\rho$ 
due to logarithmic scaling in bitrate. 
These 
experiments demonstrate substantial savings even at modest $\rho = 2.8$.

\begin{Rem}
This comparison provides context rather than a strict benchmark, as single- and multi-channel architectures embody different tradeoffs: single-channel schemes rely on oversampling and complex recovery for hardware compactness, whereas ECRT leverages channel diversity for Nyquist-rate sampling with $\mathcal{O}(N)$ CRT-based reconstruction at the cost of additional front-end circuitry.
\end{Rem}

\section{Conclusion}
\label{sec:con}
This paper presented an ECRT scheme for two-channel modulo ADCs that transmits one
quantised channel together with a scaled difference, thereby removing the redundancy
of RCRT while preserving robustness to folding and quantisation errors. The analysis shows
that the required bitrate grows only logarithmically with $\rho$, and hardware results
demonstrate comparable robustness at substantially lower bitrates. These features make ECRT
well suited for low-bitrate, high-fidelity acquisition in resource-limited sensing systems.

\newpage
\balance
\bibliographystyle{IEEEbib}
\bibliography{refs}

\end{document}